\documentclass[%
 reprint,
 amsmath,amssymb,
 aps,
]{revtex4-2}

\usepackage{graphicx}% Include figure files
\usepackage{dcolumn}% Align table columns on decimal point
\usepackage{bm}% bold math
\usepackage{physics}
\usepackage[version=4]{mhchem}
\usepackage{xcolor}
\begin{document}
\def\mB{{\mathcal{B}}}
\def\mE{{\mathcal{E}}}
\def\mA{{\mathcal{A}}}
\def\br{{\mathbf{r}}}

%\preprint{APS/123-QED}

\title{
Energy Landscape Design Principle for Optimal Energy Harnessing by Catalytic Molecular Machines
}

\author{Zhongmin Zhang}
\author{Vincent Du}
\author{Zhiyue Lu}
\email{zhiyuelu@unc.edu}
\address{Department of Chemistry,
University of North Carolina,
Chapel Hill, NC 27599-3290, U.S.A.}

\begin{abstract}
Under temperature oscillation, cyclic molecular machines such as catalysts and enzymes could harness energy from the oscillatory bath and use it to drive other processes.
Using a novel geometrical approach, under fast temperature oscillation, we derive a general design principle for obtaining the optimal catalytic energy landscape that can harness energy from a temperature-oscillatory bath and use it to invert a spontaneous reaction. 
By driving the reaction against the spontaneous direction, the catalysts convert low free energy product molecules to high free energy reactant molecules.
The design principle, derived for arbitrary cyclic catalysts, is expressed as a simple quadratic objective function that only depends on the reaction activation energies, and is independent of the temperature protocol. Since the reaction activation energies are directly accessible by experimental measurements, the objective function can be directly used to guide the search for optimal energy-harvesting catalysts. 
\end{abstract}

\maketitle

In stochastic thermodynamics, catalysts and enzymes can be considered cyclic molecular machines \cite{westerhoff1986enzymes, rozenbaum2004catalytic, qian2005cycle, astumian2001making, yasuda2001resolution,hill1975stochastics, qian1997simple, qian2000simple, astumian2002brownian, reimann2002brownian}: the catalyst undergoes a cycle of state changes to assist the conversion of reactant(s) to product(s), and returns to its initial state. In a stationary environment, the catalytic cycle reaches a nonequilibrium steady state (NESS), driven by the thermodynamically spontaneous reaction ($\Delta G< 0$) to a biased direction.

In idealized stationary environments, molecular machines can transduce free energy from one form to another \cite{hill2013free, chen1987asymmetry, tsong1988electroconformational}. By contrast, molecular machines in realistic time-varying environments can demonstrate novel dynamical and thermodynamic behavior beyond NESS.
For example, a periodically oscillating environment could drive a detailed-balanced system to mimic a dissipative system \cite{PhysRevX.6.021022,busiello2018similarities}.
Moreover, the driving force provided by the time-changing environment could drive enzymes or molecular complexes to function as engines, ratchets, or pumps \cite{astumian1994fluctuation, horowitz2009exact, PhysRevLett.109.203006, astumian2007design, rahav2008directed, sinitsyn2007berry,astumian2007adiabatic,astumian2003adiabatic,astumian2001towards,parrondo1998reversible, astumian2018stochastically,Li1997-xy,reimann1996brownian}. Also, periodically oscillating temperatures could drive catalysts to alter the reaction kinetics or even shift the equilibrium concentration \cite{berthoumieux2007response,Lemarchand2012-jl, lemarchand2012chemical, berthoumieux2009resonant}. These results indicate that catalysis could also demonstrate novel behavior in time-varying environments. 

Many existing works on molecular ratchet focus on their dynamics with a given fixed energy landscape \cite{derenyi1999generalized, astumian2007adiabatic, derenyi1999efficiency,astumian2007adiabatic,ai2005heat, reimann2002brownian, seifert2012stochastic, parrondo2002energetics, sekimoto2000carnot, hondou2000unattainability, asfaw2004current, van2007carnot,Li1997-xy,reimann1996brownian}. 
However, only recently people started to explore the \emph{optimal design of the energy landscape} for functional molecular ratchets and pumps \cite{brown2017toward,lucero2019optimal,brown2019theory, brown2018allocating}. This work focuses on identifying a novel regime of driven catalysis and the corresponding design principles of its optimal energy landscape.

Consider a catalyst and a chemical reaction whose forward direction is always spontaneous for a continuous range of stationary temperatures. If temperature oscillates within the range, can the catalyst drive the reaction backward? If yes, the catalyst harnesses environmental energy to convert low free energy products to high free energy reactants. Although counter-intuitive, this effect is similar to the Parrondo's Paradox \cite{harmer1999losing}, where a gambler can win a game by periodically switching between two losing strategies.

\begin{figure}
    \centering
    \includegraphics{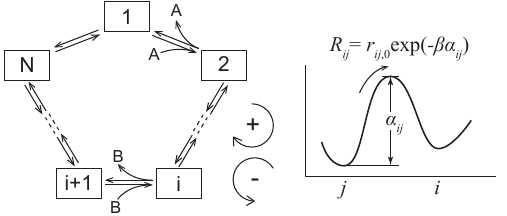}
    \caption{A general model of a catalyzed reaction, where the catalyst undergoes a cyclic pathway consisting of $N$ intermediate states and $2N$ transitions. A and B represents the sets of reactants and products, which can join/leave the catalytic loop at arbitrary locations. We choose a convention that the forward reaction goes clockwise. The transition from $j$ to $i$ and its rate $R_{ij} $is illustrated by the energy landscape.}
    \label{Fig:Markov}
\end{figure}

This letter derives a universal objective function to find catalyst's energy landscape that can maximally drive the reaction against its spontaneous direction. The objective function (Eq.~\ref{eq:CCC}) is simply related only to the activation energies in the catalytic cycle, shown as $\alpha_{ij}$ in Fig.~\ref{Fig:Markov}. Thus, this theory is directly applicable to experimental selections of catalysts or designing catalytic reaction pathways to achieve energy harvesting.  

Consider a general Markov model of the cyclic kinetics of catalysis sketched in Fig.~\ref{Fig:Markov}, as a single-loop catalytic pathway consisting of N states. By completing a cycle, the reactant A is converted into product B, and the catalyst returns to its initial state. There are $2N$ transitions on the $N$-state cycle between adjacent states whose rates follow the Arrhenius law, 
\begin{equation}
\label{eq:Arrh}
    R_{ij} = r_{ij,0}\exp(-\beta \alpha_{ij})
\end{equation}
for state $j$ to $i$. Here $\beta$ is the inverse temperature, and $\alpha_{ij}$ is the activation energy of the transition from $j$ to $i$
(see Fig.~\ref{Fig:Markov}). The temperature-independent prefactor, $r_{ij,0}$, can be proportional to the concentration of external molecule (A or B) if an external molecule is absorbed in the transition.

The dynamics of the catalysis can be described by the master equation
\begin{equation}
    \dv{\vec{p}}{t} = \hat{R}(\beta) \cdot \vec{p}
\end{equation}
where $\vec{p}$ is a $N$-dim column vector characterizing the probability of each state, $\hat R(\beta)$ is the transition rate matrix at given inverse temperature $\beta$, the off-diagonal elements of $\hat {R}$ is $R_{ij}$, and the diagonal elements are chosen such that each column of $\hat{R}$ sums to $0$. 

Throughout this paper, we assume that the chemical bath is infinitely large, so the chemical concentrations remain constant. Then at a fixed temperature, this system reaches a NESS, $\vec p^{\,ss}$ where $\hat{R}(\beta) \cdot \vec{p}^{\,ss}=0$. Then spontaneous reaction's rate is characterized by the net NESS probability current:
\begin{equation}
    \label{eq:nessj}
    J^{ss} = {R}_{21}  p^{ss}_{1}-{R}_{12}  p^{ss}_{2}
\end{equation}
where at NESS, the current is uniform across the loop, and we choose to define it between states $1$ and $2$. In this work, we choose the convention that clockwise (CW) current is positive, corresponding to the forward reaction (A to B), and the counter-clockwise (CCW) current is negative. If the Gibbs free energy of A is higher than B ($G_A>G_B$), the forward reaction is spontaneous, leading to a positive NESS current $J^{ss}>0$. The affinity, which is equal to the free energy difference between A and B, determines the direction:
\begin{equation}
    \mA \equiv \beta^{-1}\sum_{\langle i,j\rangle} \varepsilon_{ij} \log R_{ij} =  G_A-G_B=-\Delta G
\end{equation}
where $\Delta G$ is the free energy change corresponding to the forward reaction, the $\sum_{\langle i,j\rangle}$ sums over all of the $2N$ transitions from $j$ to $i$ on the loop. We have adopted the following sign indicator of transition from $j$ to $i$:
\begin{equation}
    \varepsilon_{ij} = 
    \left\{
    \begin{array}{cl}
         1& \quad\mathrm{forward }  \\
         -1& \quad\mathrm{backward }
    \end{array}
    \right.
\end{equation}
If $\mA>0$, the spontaneous reaction is forward, $\mA<0$ backward, and if $\mA=0$, the system is at thermal equilibrium without net reaction flow. In this paper, we assume $G_A$ is always greater than $G_B$ within the temperature range of interest, and thus $\mA>0$ and $J^{ss}>0$, and the spontaneous reaction always goes forward (CW).

If the temperature non-quasi-statically oscillates in time, the chemical reaction is driven out of NESS. The system eventually reaches a time-periodic state (i.e., a periodic orbit in probability space):  $\vec p(t+\tau)=\vec p(t)$, where $\tau$ is period of temperature oscillation. There have been studies of the periodic states under periodic temperature modulation \cite{berthoumieux2007response,Lemarchand2012-jl, lemarchand2012chemical}. 
However, it is generally impossible to analytically solve the dynamics for arbitrary systems or arbitrary reaction landscapes. To derive the generic design principle, in this letter, we consider the \emph{fast oscillation limit} $\tau \rightarrow 0$ \footnote{Although the optimal design principle is derived under the fast oscillation limit, Fig.~3 indicates that the optimal catalyst here is able to invert reaction direction even under finite driving frequencies. }, where a perturbation analysis \cite{tagliazucchi2014dissipative} for small periods $\tau$ could reveal an analytical solution of the periodic orbit shrinking into a fixed point $\vec {p}(t) \rightarrow \vec p^{\,*}$, which leads to a general principle that applies to arbitrary reaction energy landscapes. (See Appendix.) The fixed point $\vec p^{\,*}$ can be considered as an \emph{effective NESS} corresponding to an \emph{effective rate matrix} $\hat R^{*}$:
\begin{equation}
    \hat R^{*} \cdot \vec p^{\,*}  = 0
\end{equation}
where the effective rate matrix is nothing but the time average of $\hat R(\beta(t))$ over a period:
\begin{equation}
    \hat R^{*} \equiv \lim_{\tau\rightarrow 0}\frac{1}{\tau} \int_{0}^{\tau} \hat R(\beta (t)) \dd t
\end{equation}
At the fast oscillation limit, one can find the average current by using $\vec p^{~*}$ and $\hat R^{*}$ similar to that in Eq.~\ref{eq:nessj}:
\begin{equation}
\label{eq:*J}
    J^* = {R^{*}}_{21}  p^{*}_{1}-{R^{*}}_{12}  p^{*}_2
\end{equation}
In contrast to NESS, the affinity is no longer well-defined since the temperature is no longer a fixed constant. Here without a constant temperature, we introduce an \emph{dimensionless affinity},
\begin{equation}
\label{eq:A'}
    \tilde \mA = \sum_{\langle i,j\rangle} \varepsilon_{ij} \log R_{ij}
\end{equation}
where $R_{ij}$ is not restricted to a fixed-temperature rate matrix $\hat R(\beta)$ but can also be defined for effective rate matrix $\hat R^{*}$. 
At a constant temperature, the $\tilde \mA$ calculated from stationary temperature rate matrix $\hat R(\beta)$ can be related back to $\mA$ by $\tilde \mA= \beta \mA(\beta)= -\beta \Delta G$. When temperature rapidly oscillates, one can define the effective dimensionless affinity $\tilde \mA^*$ by plugging the effective rate matrix $\hat R^*$ in Eq.~\ref{eq:A'}.

Under oscillatory temperature, the direction of reaction (the sign of $J^*$) is solely determined by the active driving force $\tilde \mA^*$: if $\tilde \mA^*>0$, the reaction on average proceeds forward ($J^*>0$); if $\tilde \mA^*<0$, the reaction on average proceeds backward ($J^*<0$). 

Thus, the goal of searching for a catalyst to invert a spontaneous reaction is formulated in terms of $\tilde \mA^*$: Consider a spontaneous reaction where $\Delta G <0$, $J(\beta)>0$, and $\tilde \mA(\beta)>0$ for any temperature within a continuous range. When temperature oscillates within the range, what catalyst facilitates a negative $\tilde \mA^*<0$ (i.e., reaction is inverted and $J^* <0$)?

\begin{figure}
    \centering
    \includegraphics{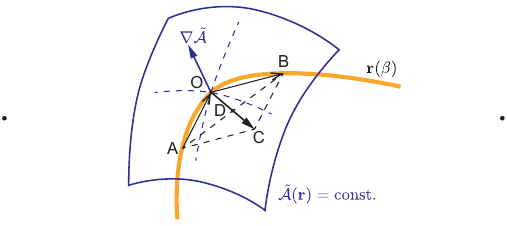}
    \caption{Illustrated is the 2N-dimensional design space of $\mathbf{r}=(R_{21}, R_{12}, ..., R_{ji}, R_{ij}, ..., R_{1N}, R_{N1})$. 
    The yellow $\beta$-locus crossing 3 points AOB is the set of $\mathbf{r}(\beta)$ corresponding to a given energy landscape at all possible inverse temperature $\beta$'s.
    %(yellow locus containing AOB and crossing manifold at point O). 
    Blue surface represents a $(2N-1)$-manifold defined by $\tilde \mA(\br)=\text{const}$ that contains point O. 
    The gradient of $\tilde \mA(\br)$ at point O, is shown as the blue normal vector $\nabla \tilde \mA$.
    When we consider temperature oscillation between $\beta_{1,2}=\beta_0 \pm \Delta \beta$, their corresponding rate matrices are points A and B. The rate matrix at $\beta_0$ is represented by point O. The effective rate matrix $\hat R^*$ corresponds to the midpoint D between A and B. At infinitesimal temperature amplitude, the vector $\protect\overrightarrow{\textrm{OC}} = \protect\overrightarrow{\textrm{OB}}-\protect\overrightarrow{\textrm{AO}}$ becomes $\dd[2]{\mathbf{r}}/\dd{\beta}^2$ in Eq.~\ref{eq:acc}. 
    }
    \label{fig:geo}
\end{figure}

In this letter, based on the geometric property of $\tilde{\mA}$, we obtain a universal objective function, Eq.~\ref{eq:CCC}, to find the optimal catalytic reaction inversion.
Historically, geometry has played important roles in thermodynamics. Gibbs first used geometry to demonstrate the thermodynamic properties within the space of state functions \cite{Weinhold1976-ia}. Recently, Crooks \cite{Sivak2012-yt}, Ito \cite{Ito2018-ci,Yoshimura2021-ie}, and Dong \cite{Li2022-ib} have derived various general thermodynamic results by utilizing differential geometry within various types of probability-distribution spaces.

Rather than working within a probability space, this letter focuses on the geometry in the $2N$-dimensional space consisting of kinetic rates: $\mathbf{r}=(R_{21}, R_{12}, ..., R_{ji}, R_{ij}, ..., R_{1N}, R_{N1})$. According to Eq.~\ref{eq:Arrh}, the reaction rates are determined by both temperature $\beta^{-1}$ and the catalytic energy landscape $\alpha_{ij}$'s. 
For any catalyst specified by its energy landscape $\alpha_{ij}$'s, its kinetic rates $\mathbf r(\beta)$ parameterized by inverse temperature $\beta$ is illustrated by a yellow $\beta$-locus in Fig.~\ref{fig:geo}. 

Without losing generality, let us illustrate our theory using a simple square-wave temperature oscillation between $\beta_1$ and $\beta_2$ of equal time duration. At constant temperature, $\beta_1$ (or $\beta_2$), the kinetic rate matrix $\hat R(\beta)$ or equivalently $\mathbf r(\beta)$ is illustrated by the point A (or B) on the $\beta$-locus in Fig.~\ref{fig:geo}. Under fast temperature oscillation, the effective rate matrix is simply the arithmetic mean:
\begin{equation}
    \hat R^* = \frac{\hat R(\beta_1)+\hat R(\beta_2)}{2}
\end{equation}
and is represented by point D, the midpoint between A and B in the $\mathbf r$ space. 

Recall that the generalized dimensionless affinity $\tilde \mA(\mathbf r)$ is a function on the $\mathbf r$-space, and its sign dictates the direction of averaged flow. By construction, within the range of $\beta \in [\beta_1,\beta_2]$, the reaction free energy $\Delta G(\beta)<0$ and $\tilde \mA(\mathbf r(\beta))>0$. Typically point D is not on the yellow locus $\mathbf r(\beta)$. Thus $\tilde \mA(\mathbf r)$ at point D can take a very different value than that on the yellow locus. When $\tilde \mA(\mathbf r)<0$ at point D, the catalyst inverts the reaction direction when temperature oscillates rapidly.  

Geometrically, the catalyst is represented by a locus $\mathbf {r}(\beta)$ in the $\mathbf r$-space. The analysis above allows us to characterize the catalyst's ability to invert reaction by how much the locus $\mathbf {r}(\beta)$ curves toward the steepest descent direction (gradient) of $\tilde \mA(\mathbf {r})$. Notice this geometrical characterization is not dependent on the specific protocol of temperature oscillation. 

The qualitative geometrical argument above can be quantified by two vectors. Firstly, the bending of the $\beta$-locus $\br(\beta)$ can be characterized by the second order derivative vector $\dd[2]{\mathbf{r}}/\dd{\beta}^2$ (see vector $\overrightarrow{\rm OC}$ in Fig.~\ref{fig:geo}). This curvature-like vector is the acceleration vector for the motion of point $\br(\beta)$ as the $\beta$ varies. 
The entries of $\dd[2]{\mathbf{r}}/\dd{\beta}^2$ are 
\begin{gather}
\label{eq:acc}
    \dv[2]{R_{ij}}{\beta} = \alpha_{ij}^2 R_{ij}
\end{gather} 
Secondly, the variation of $\tilde \mA$ in the $\br$-space is characterized by the gradient vector $\nabla\tilde \mA(\mathbf {r})$ (as the blue arrow in Fig.~\ref{fig:geo}), whose entries are
\begin{equation}
\label{eq:grad}
   \pdv{\tilde \mA}{R_{ij}} = \frac{\varepsilon_{ij}}{R_{ij}}~.
\end{equation}
Combining the above, the catalyst's ability to invert the reaction direction is characterized by the inner product between the second-order derivative vector $\dd[2]{\mathbf{r}}/\dd{\beta}^2$ and the gradient vector of $\tilde \mA$:
\begin{equation}
\label{eq:CCC}
    \mathcal{C}(\{\alpha_{ij}\}) = \nabla \tilde \mA \cdot \dv[2]{\mathbf{r}}{\beta} = \sum_{\langle i,j\rangle} \varepsilon_{ij} \alpha_{ij}^2 
\end{equation}
which serves as a universal objective function to find the optimal catalytic energy landscape that can achieve strong reaction inversion. 

An alternative derivation based on a finite-difference analysis of temperature oscillation between $\beta_0-\Delta \beta$ and $\beta_0+\Delta \beta$ is shown in Fig.~\ref{fig:geo} and the SI. Here $\mathcal{C}(\{\alpha_{ij}\})$ is directly proportional to $\Delta \tilde \mA$, the difference of $\tilde \mA$ of $\hat R^*$ and the NESS $\tilde \mA(\beta_0)$ at constant temperature $\beta_0$. 
\begin{equation}
\label{eq:expand}
    \Delta \tilde \mA\equiv \tilde \mA(\hat R^*) - \tilde \mA(\mathbf r(\beta_0)) =\mathcal{C}(\{\alpha_{ij}\}) \frac{\Delta \beta^2}{2}  +o(\Delta \beta^2)
\end{equation}

Due to the nice geometric property of the constant-$\tilde \mA(\br)$ manifold and the $\beta$-locus \footnote{In the logarithm $\br$-space, the constant-$\tilde \mA$ manifolds become a set of parallel hyper-planes with unidirectional gradients, and the $\beta$-locus for any energy landscape is a straight line. The gradient of $\tilde \mA$ and the acceleration vector of the $\beta$-locus has an elementary-wise inverses dependence on $R_{ij}$.}, the $R_{ij}$ from Eqs.~\ref{eq:acc} and \ref{eq:grad} cancels out, and the resulting objective function Eq.~\ref{eq:CCC} takes a simple quadratic form that only depends on the activation energies of the catalyst $\alpha_{ij}$'s, and is independent of the specific temperature protocol. 
For the same geometrical reason, $\mathcal C$ applies to large-amplitude temperature oscillation (see Fig.~\ref{fig:scatter}b).

The objective function $\mathcal{C}(\{\alpha_{ij}\})$ is directly accessible by experiment via direct measurements of the activation energies $\alpha_{ij}$'s. Thus, our result (Eq.~\ref{eq:CCC}) provides chemists with an easy approach to predict arbitrary catalysts' ability to invert reaction direction under fast temperature oscillation.

\begin{figure}
    \centering
    \includegraphics{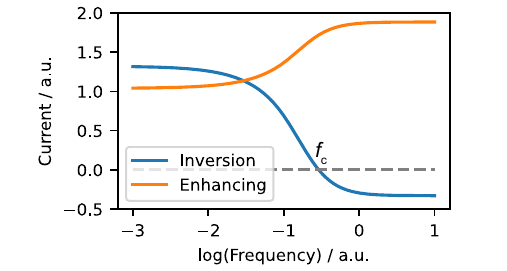}
    \caption{Reaction rate (average probability current at periodic steady state) versus temperature oscillation frequency $f=\tau ^{-1}$ for both the optimal energy landscapes for reaction inversion and enhancement, obtained for $\mA=-\Delta G=1$, $\alpha_{max}=11$, $\beta_0=0.9$ and $\Delta \beta=0.3$.}
    \label{fig:freq}
\end{figure}

For illustration, Fig.~\ref{fig:freq} demonstrates the frequency response of the optimal 3-state catalyst landscape ($\{\alpha_{ij}\}$) under the following design constraints. First, we fix the reaction's free energy change $\Delta G(\beta)=-1$ for all $\beta$'s. As a result, the affinity must be constant $\mA=-\Delta G=1$ regardless of the choice of the catalyst. Secondly, we assume that the reaction rate prefactor $r_{ij,0}$'s are all fixed to be the same constant. By doing so for any constant temperature,
\begin{equation}
\label{eq:A-plane}
    \mA = - \sum_+ \alpha_{ij} + \sum_- \alpha_{ij} = 1
\end{equation}
where $\sum_+$ (or $\sum_+$)  is the sum over all forward (or backward) reaction transitions. 
Thirdly, as we search for the optimal landscape (activation energies), we restrict the activation energies of all $2N$ transitions in the range $\alpha_{ij}\in[0,\alpha_{max}]$. Here we assume the reaction's affinity is weaker than the maximum allowed activation energy: $\mA=1<\alpha_{max}$.

According to our result, to find the optimal catalyst with the strongest driving force against spontaneous reaction under temperature oscillation, one needs to minimize the objective function 
\begin{equation}
    \mathcal C=\sum_+ \alpha_{ij}^2 - \sum_- \alpha_{ij}^2
\end{equation}
which is a simple quadratic minimization problem on a convex set. 
For $N=3$, the optimal solution of $\alpha_{ij}$'s are
\begin{align}
    \label{eq:+inv}
    \alpha^\text{inv}_+ &= (\frac{2\alpha_{max}-1}{3}, \frac{2\alpha_{max}-1}{3}, \frac{2\alpha_{max}-1}{3}) \\
    \alpha^\text{inv}_- &= (\alpha_{max}, \alpha_{max}, 0)
    \label{eq:-inv}
\end{align}
where $\alpha^\text{inv}_+$ are the activation energies $\alpha_{ij}$'s for the 3 forward transitions and $\alpha^\text{inv}_-$ are for the 3 backward transitions (see SI).  The order of the 3 forward (or 3 backward) $\alpha_{ij}$'s does not impact the result. 

Beyond the fast oscillation limit, the optimal catalyst can invert the reaction at finite frequency $f$'s (see Fig.~\ref{fig:freq}). At the critical frequency $f=f_c$, the reaction free energy force $\Delta G$ is completely stalled by inversion force from the catalyst, and the reaction stops ($J_\textrm{period}=0$); at larger frequency, $f>f_c$ the catalyst's driving wins over $\Delta G$ and reaction direction is inverted ($J_\textrm{period}<0$). 
At the fast oscillation limit $f \gg 1$, we find that $J^*= (R^*_{13}R^*_{32}R^*_{21}-R^*_{12}R^*_{23}R^*_{31})/\kappa$, where $\kappa>0$ (see SI). Thus the sign of current $J^*$ is always the same with $\tilde \mA^*$.

It is worth pointing out that the objective function Eq.~\ref{eq:CCC} can be used toward an inverse effect of catalytic reaction inversion,
i.e., driving force enhancing.
%In the inverse effect, temperature oscillation can drive a catalyst to enhance the thermodynamic driving force of a spontaneous reaction, rather than inverting it. To find the optimal catalyst for reaction enhancement, one need to make the effective $\tilde \mA^*$ larger than the NESS $\tilde \mA$, by maximizing Eq.~\ref{eq:CCC} (rather than minimizing it in the reaction inversion problem). 
By maximizing Eq.~\ref{eq:CCC}, (see SI)
%The restricted maximization of $\mathcal C$ is achieved at (see SI):
\begin{align}
    \label{eq:+enh}
    \alpha^\text{enh}_+ &=(\alpha_{max}, \alpha_{max}, 0) \\
    \alpha^\text{enh}_- &= (\frac{2\alpha_{max}+1}{3}, \frac{2\alpha_{max}+1}{3}, \frac{2\alpha_{max}+1}{3})
    \label{eq:-enh}
\end{align}
defines a reaction-enhancing catalyst that optimally enhances the spontaneity of a reaction. The reaction current enhanced at various frequencies $f$ is shown in Fig.~\ref{fig:freq}.

Even though $\mathcal C$ (Eq.~\ref{eq:CCC}) is obtained from the local curvature, due to nice geometric properties of $\tilde \mA(\br)$ and $\br(\beta)$, it remains a good optimization objective function even for big-amplitude temperature oscillations (e.g., for $\beta_0=0.9$, $\Delta \beta=0.3$). 
We demonstrate that for both small and large amplitudes, $\mathcal C$ is approximately linearly correlated to the change of thermodynamic driving force. The linear correlation is shown in Fig.~\ref{fig:scatter} for both small $\Delta \beta =0.05$ and large $\Delta \beta=0.3$ by scatter plots of $10^4$ points. Each point is obtained from one randomly generated energy landscape $\{\alpha_{ij}\}$. Notice in the small amplitude limit, $\Delta\beta\ll1$, $\mathcal C$ is equal to $2\Delta \tilde \mA/\Delta\beta ^2$ (Eq.~\ref{eq:expand}). The optimal catalysts for reaction inversion and enhancement (obtained by minimizing and optimizing $\mathcal C$) are highlighted as red and black crosses, appearing at the two ends of both scatter plots.

\begin{figure}
    \centering
    \includegraphics{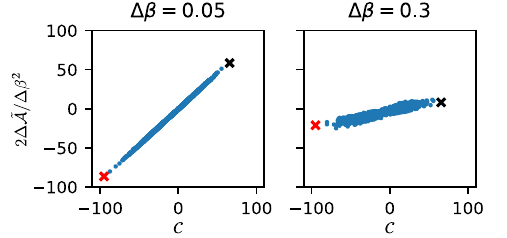}
    \caption{Objective function $\mathcal C$ and the scaled affinity change $2\Delta \tilde \mA/\Delta\beta ^2$ for randomly generated energy landscapes under the same set restriction with the optimization problem ($\mA=-\Delta G=1$, $\alpha_{max}=11$, $\beta_0=0.9$). The red and black crosses correspond to the optimal energy landscapes for reaction inversion (minimizing $\mathcal C$) and enhancement (maximizing $\mathcal C$).}
    \label{fig:scatter}
\end{figure}

At stationary temperature, kinetic intuition may argue that the higher the activation energy, the slower the corresponding transition rates. However, when temperature oscillates, our theory indicates that big variation in the activation energies of the inverse reaction direction and mild activation energies of the forward reaction direction could suppress the forward reaction and favor the inverse direction. 
Moreover, designing the catalytic reaction inversion suffers from a trade-off relation between strength and speed. Strong inversion (large $|\Delta \tilde \mA|$) favors the choice of larger activation energies (larger $\alpha_{ij}$), which impedes the net reaction current.

In conclusion, this letter demonstrated a geometric approach to derive the general design principle of optimal oscillatory-driven catalysis. In this regime, we demonstrate catalysts that can harness energy from an oscillatory-temperature bath and utilize the energy to enhance or invert a spontaneous reaction.
The design principle is formulated by an objective function Eq.~\ref{eq:CCC}, which only depends on the activation energies of the energy landscape in a quadratic form. Due to the nice geometric property of the thermodynamic driving force $\tilde \mA$, the objective function is independent of the temperature protocol. Moreover, this result obtained from the fast oscillation limit can still be used to invert spontaneous reactions at finite-frequency temperature oscillation. 
Since activation energies are accessible in the experimental study of reaction mechanisms, this result could be experimentally verified and directly used to guide the design of useful catalysts for energy harnessing.

\begin{acknowledgements}
We acknowledge the financial support from the startup fund at UNC-Chapel Hill and the fund from the National Science Foundation Grant DMR-2145256. Z.L. appreciates helpful discussions with Hong Qian and suggestions on the manuscript from Chase Slowey. We also appreciate the reviewers' comments to help enhance the clarity of the presentation.
\end{acknowledgements}

\bibliography{ref}% Produces the bibliography via BibTeX.

\end{document}